# Stepwise Computational Synthesis of Fullerene $C_{60}$ derivatives.
# 2. Hydrogenated Fullerenes from $C_{60}$ to $C_{60}H_{60}$

Elena F.Sheka

*Research Department, Peoples` Friendship University of the Russian Federation, Moscow*
sheka@icp.ac.ru

**Abstract.** Hydrogenation from $C_{60}$ to $C_{60}H_{60}$ has been studied by unrestricted broken spin symmetry Hartree-Fock approach implemented in semiempirical codes based on AM1 technique. The calculations were focused on sequential addition of hydrogen molecules to the fullerene cage following the indication of the cage target atoms by the highest atomic chemical susceptibility calculated at each step. The obtained results are analyzed from energetic, symmetry, and the composition abundance viewpoints.

## 1. Introduction

The effectively-non-paired-electron concept of the fullerene atoms chemical susceptibility [1-3] lays the foundation of the stepwise computational synthesis of the hydrogenated derivatives (hydrogenates below) of fullerene $C_{60}$ performed in the current study. Firstly sampled by the fluorination from $C_{60}$ to $C_{60}F_{60}$ (see Part 1 [4]), the methodology was applied to hydrogenation from $C_{60}$ to $C_{60}H_{60}$ by using unrestricted broken symmetry HF (UBS HF) SCF semiempirical approach (UBS HF version of the AM1 technique of the CLUSTER-Z1 codes). The calculations were focused on sequential addition of hydrogen molecules to the fullerene cage. A complete family of species $C_{60}H_{2k}$ $k$=1,...,30 has been produced. Avoiding the many-fold isomerism problem that was discussed in details in Part 1 and that consists in enormous number of isomers for, say, $C_{60}H_{36}$ (600 873 146 368 170) and $C_{60}H_{48}$ (23 322 797 475) [5], the preferred binding sites for sequential additions are selected by the largest value of atomic chemical susceptibility (ACS) quantified by the effectively non paired electron fraction $N_{DA}$ on the considered atom $A$ [1, 2]. As shown, any addition of hydrogen molecule causes quite noticeable change in the $N_{DA}$ distribution over the $C_{60}$ cage atoms. That is why the synthesis was performed as a series of predicted sequential steps. The reaction starts around a pair of fullerene $C_{60}$ atoms with the biggest $N_{DA}$ values. The atoms usually form one of short-length bonds within one of six identical naphthalene-core fragments. When the first adduct $C_{60}H_2$ is formed, the reaction proceeds around a new pair of its fullerene cage atoms with the highest $N_{DA}$ values resulting in the formation of $C_{60}H_4$ adduct. A new $N_{DA}$ map reveals the sites for the next addition step and so on. Following these methodology, 30 steps of the $C_{60}$ to $C_{60}H_{60}$ hydrogenation were made. The obtained results are analyzed from the energetic, symmetry, and the composition abundance viewpoints and are compared with experimental data where possible.

## 2. Grounds of computational methodology

Hydrogenation of $C_{60}$ was one of the first reactions related to fullerenes that were considered computationally. The list of available publications is rather long (see a detailed review of papers published by the beginning 1994 in [6], later papers [7-9], and a collection of works by Clare & Kepert [10-16]). Altogether, the calculations exhibit practically all available computational schemes applied to the species. But not the dispersion of the results caused by different techniques was the reason of so numerous investigations. The main point concerned structural models of hydrogenates in view of the above-mentioned many-fold isomerism problem. Experimental findings that could help to solve the problem are rather scarce until now and come to the following: 1) polyaddition derivatives are abundant among the products formed; 2) the products are characterized by even number of hydrogen atoms; 3) isomerism appearance is limited to a single isomer, rarely to two-three species [17-19]. Therefore, to be practically useful simulations had to deal with the construction of one and/or a few properly selected isomers of polyaddition hydrogenates. The only clear evidence how one should reach the goal is concerned with the pair-hydrogen-atom addition at each subsequent step. This was common for hydrogenation of both kinds, namely, for molecular one that is characteristic for the Birch reduction of $C_{60}$ [17] and for atomic-radical ones that occurs either under transfer hydrogenation of $C_{60}$ with dihydroanthracene [20] or under hydrogenation of fullerite at elevated temperatures and hydrogen pressures [21].

When constructing the structural models needed for calculations the first question arises how a single pair of hydrogen atoms is accommodated on the fullerene cage implying either 1,2- (suitable for the addition of both hydrogen molecule and a pair of separated hydrogen atoms) or 1,4- addition (suitable only for separated atoms). The differences in the energy was attributed to the result that only 1,2-addition does not disrupt the remaining polyene structure of the cage, whereas one double (short) bond is moved to a five-membered ring in the case of 1,4-addition. A thorough study [22] showed the preference of 1,2-addition for hydrogen and made it possible to consider subsequent steps as one-by-one addition of the hydrogen molecule to the fullerene cage.

However, there have still been two other questions related to understanding how the addition of each next hydrogen pair occurs and what reasons govern the selection of a single isomer. As for the manner of the successive addition, a contiguous conjectured route to the formation of polyhydrogenates was accepted [9, 18]. As for the isomer selection, the main suggestion concerned isomers of high symmetry followed by further selection governed by favoring a single one with the least total energy. Originated at the first experimental producing $C_{60}H_{36}$ where $T_h$ symmetry of the species was proposed [17], the high-symmetry selection was supported by structural characterization of $C_{60}H_{18}$ [18] for which a $C_{3v}$ symmetry crown structure was reliably established. Afterwards the idea became common for all calculations related to polyhydrogenates, particularly for $C_{60}H_{36}$ [7, 8, 12-16].

Common to all performed computations is using close-shell restricted computational schemes. Applied to the $C_{60}$ fullerene cage it means accepting a complete covalent bonding of the molecule odd electrons. The latter are considered as classic $\pi$ electrons typical for the benzene molecule. However, as will be shown below, this approach is valid only for high hydrogenates with $k \geq 20$ when the effectively unpaired odd electron pool characterized by the molecular chemical susceptibility (MCS) $N_D$ is practically worked out. For lower hydrogenates, particularly for the lowest ones at $k \leq 10$, real energies of the species are significantly lower than those obtained in the close-shell approximation that makes the restricted solutions unstable. Oppositely to the case, the UBS HF approach used in the current study provides computations of the species with much lower energy and thus more stable. Additionally it reveals the effectively unpaired odd electron pool and offers MCS $N_D$ as a generalized measure of the reaction ability.

UBS HF approach similarly to the restricted HF one used in [22] reveals a preference of the 1,2-addition against 1,4-one. In what follows, this allowed for considering all sequential steps of hydrogenation as 1,2-addition stepwise implementations. All hydrogen atoms are bound to the exterior surface of the cage. As for both the contiguous conjectured route and high symmetry criterion for the isomer selection, our approach does not need any of these assumptions since changing ACS $N_{DA}$ map after each addition performed creates a definite algorithm of the $H_2$ addition in accordance with the highest $N_{DA}$ values. Moreover, the two assumptions validity can readily be checked in due course of the computational synthesis performed.

## 3. Algorithm of computational synthesis of $C_{60}$-hydrogenates

Analysis of the odd-electron enhanced chemical activity of the $C_{60}$ molecule highlights its chemical portrait presented by the ACS ($N_{DA}$) distribution over the molecule atoms [1-3]. According to the latter, the molecule consists of six identical naphthalene-core fragments forming a $6*C_{10}$ configuration. Using different colors for atoms of different ACS, Fig.1 visualizes the portrait in the form of either molecular structure (Fig.1a) or its projection on Schlegel diagram (Fig.1b) whereas the corresponding numerical ACS distribution is given in Fig.1c. Coloring on molecular structure and numeration on Schlegel diagram correspond to the atom numeration in the output file. The ACS values in Fig.1c are plotted from the highest to the lowest one in the A→Z manner. The plotting reveals five groups of atoms that consist of six identical pairs and are shown in Fig.1a and Fig.1b by different colors.

According to the figure, the initial step of any addition reaction must involve atoms of group 1 characterized by the highest ACS. One may choose any of six pairs to start the reaction of attaching one hydrogen molecule to the fullerene cage as well as any other either atomic or molecular addend. When the first adduct $C_{60}H_2$ is formed, the reaction proceeds around the new cage atoms with the highest $N_{DA}$ values resulting in the formation of adduct $C_{60}H_4$. A new ACS map reveals the sites for the next addition step and so on. Following these methodology, a complete list of fluorinated fullerenes $C_{60}F_{2k}$ has been synthesized in Part 1. Repeating the procedure described in details elsewhere [4], a complete list of $C_{60}$-hydrogenates was obtained in the current study.

## 4. Results and discussion

### 4.1. $C_{60}$ hydrogenation as algorithmic process

When starting hydrogenation of $C_{60}$, the hydrogen molecule is placed in the vicinity of the selected atoms of group 1 (33 and 22 in the case, see Fig.1b) and a full optimization of the complex geometry in the singlet state is performed. As occurred, the hydrogen molecule is willingly attached to the cage forming two C-H bonds, if only the molecule axis is not normal to the chosen C-C bond. In the latter case the molecule is repelled from the cage. Therefore, the reaction with molecular hydrogen occurs in one stage oppositely to two-stage molecular fluorination [4]. Chart 1 exhibits the action of processing algorithm related to $C_{60}$ to $C_{60}H_{18}$ hydrogenation, additionally illustrated by Schlegel diagrams in Fig.2.

The chart presents fragments of the ACS A→Z lists involving the highest rank data related to the corresponding hydrogenates. As follows from the chart, the ACS list of $C_{60}H_2$ (H2 for simplicity) points to two pairs of atoms, namely 5 & 3 and 60 & 57 that form two short bonds and are fully identical resulting in producing isoenergetic $C_{60}H_4$ (H4). Any other atoms differ by the ACS value considerably. As seen at the diagram, these bonds are not contiguous to the starting one similarly to the case of F4 adduct [4]. The ACS list of H4 highlights at least four high rank atoms. The first two atoms 31 and 34 are not bonded via a short bond and their bond partners (32 and 35, respectively) are located quite far in the depth of the ACS list (dotted lines in the chart indicate the availability of intervals between the data). And only atoms 40 & 54 form a short bond. Facing the situation, one has to perform a set of calculations based on 31&32, 34&35 and 40&54 additions. Obviously, the preference should be done to the isomer with the least energy. As seen from the corresponding part of the chart, the isomer analysis by the total energy definitely favors H6 (40, 54) isomer. Proceeding with H8 synthesis, one faces the same situation as previously. The high-rank-data fragment of the ACS list of H6 is headed by atoms 55 and 24 that are not bond-joined. And again a set of computations concerning 55&38 and 24&23 additions should be performed. As previously, addressing the total energy selection of isomers one gets H8 (55, 38). The same algorithm was used at each next step of synthesis so that a series of hydrogenates H10 (52, 51), H12 (31, 32), H14 (24, 23), H16 (42, 48), and at last H18 (58, 59) was obtained. When passing from H16 to H18 there was no alternative to the combination (58, 59) according to the ACS list. The obtained H18 hydrogenate is of crown structure of $C_{3v}$ symmetry which is in full consistence with experimental findings based on $^1$H NMR [18] and $^3$He NMR [29] studies.

A series of diagrams in Fig.2 visualizes the successive chain of cage atoms that participate in the synthesis in due course of $C_{60}$ to $C_{60}H_{18}$ hydrogenation. As seen from the figure, the atoms do not follow a contiguous conjectured route. And this feature proceeds when going to high hydrogenates some of which are shown in Fig.2. As said earlier, the addition of hydrogen molecule at each successive step makes the choice of 1,2-addition quite mandatory. This requirement provides keeping polyene structure of the fullerene cage that is why none of the obtained hydrogenates possesses a short bond moving to a pentagon frame. The feature excludes the availability of $T_h$, $T$, and $S_6$ isomers that are characterized by short-bond framing of pentagons among, say, $C_{60}H_{36}$ and $C_{60}H_{48}$ species. The latter hydrogenates shown in Fig.2 look highly symmetrical but their exact symmetry is $C_1$. Obviously, the deviation from high symmetry is rather minor so that the products will show high symmetry pattern in various experiments.

Geometric parameters, symmetry, and total energy of a complete series of the $C_{60}$ hydrogenates are given in Table 1. Table involves as well pairs of the cage atoms that participate in the hydrogenation at each step. The hydrogenation route is fully similar to that of fluorination [4] with the only difference in the number of the cage starting atoms: 33 & 22 for hydrogenation and 32 & 31 for fluorination. This finding exhibits actual identity of atom pairs within group 1 belonged to one of five naphthalene-core fragments. In its turn, the obtained similarity strongly supports structural parallelism between hydrogenates and fluorinates of $C_{60}$ and $C_{70}$ pointed out in numerous experimental studies (see [30] and references therein).

However, when comparing similar suits from the two families from the viewpoint of their production in practice, the parallelism does not occur fully complete. Thus, suits involving $C_{60}X_{36}$, $C_{60}X_{48}$, and $C_{60}X_{60}$ products behave quite differently: if $C_{60}F_{36}$ and $C_{60}F_{48}$ were efficiently produced and $C_{60}F_{60}$ was mentioned as a trace [24, 28], only $C_{60}H_{36}$ was obtained in a significant quantity whereas no mention about either $C_{60}H_{48}$ or $C_{60}H_{60}$ is known. A high susceptibility of $C_{60}$-hydrogenates to oxidation is one of the reasons preventing high hydrogenates from both stabilizing and efficient production. Another reason is unique to either hydrogenation or fluorination. As will be shown below, UBS HF approach made it possible to highlight this intimate difference.

As mentioned, $C_{60}H_{36}$ is the most abundant amongst other hydrogenated products. A lot of efforts were undertaken to determine the molecule structure however until now its structure has remained unsolved. $^1$H NMR technique, $^{19}$F NMR analogue of which occurred very informative in the case of fluorinated species [24-28], was inefficient in the case of hydrogenates exhibiting a broad band with few distinguishing features, consistent either with more than one isomer or with allylic oxidation during acquisition of the spectrum [19]. A spectroscopic study of $C_{60}H_{36}$ showed a complicated absorption spectrum assignment of which required a suggestion of mixing isomers [30]. $^3$H NMR study was the most successful [19]. The obtained spectra showed a complete similarity between $C_{60}H_{18}$ & $C_{60}F_{18}$ and $C_{60}H_{36}$ & $C_{60}F_{36}$ pairs of species expressed as a complete identity of the spectra structure with the only difference in the line position. When the $C_{60}H_{18}$ & $C_{60}F_{18}$ spectra are clearly evident for the $C_{3v}$ species, the structure of the $C_{60}H_{36}$ & $C_{60}F_{36}$ spectra is more complicated which was explained by the mixture of $T$ and $C_3$ isomers on the basis of calculations at the restricted Hartree-Fock level. It should be noted that starting from the first production of $C_{60}H_{36}$ the interpretation of empirical results was subordinated to looking for high symmetry isomers. This was in parallel with a fundamental assumption of computations that considered high symmetry as a governing factor in the selection of a particular single isomer.

The symmetry of the $C_{60}H_{36}$ molecule obtained in the current study is $C_1$ similarly to $C_1$ symmetry of $C_{60}F_{36}$ in [4]. Since the high-symmetry approach to the interpretation of empirical data occurred to be quite ambiguous, one can not exclude that the application of $C_1$ structure to the description of experimental spectroscopic and $^3$H NMR data will be successful enough as it was in the case of $C_1$ $C_{60}F_{48}$ Rietveld refining of high-resolution X-Ray powder diffraction structure data for $C_{60}F_{48}$ powder [4]. It should be noted that the obtained $C_1$ symmetry is a logical consequence of the hydrogenation process consideration as a successive series of 1,2-additions. Such succession conserves the cage polyene structure keeping short C-C bonds within naphthalene-core fragments. At the same time, high symmetry structures of $C_{60}H_{36}$ ($C_{60}F_{36}$) such as $T_h$, $D_{3d}$, $S_6$ are accompanied by moving a short bond to a five-membered ring [8], thus requiring 1,4-addition mode [22]. It is difficult to correlate the sufficiency of the 1,2-mode to provide $C_{3v}$ $C_{60}H_{18}$ with the necessity of its transformation into the 1,4-mode to provide high symmetry of either $C_{60}H_{36}$ or $C_{60}H_{48}$ within the same continuous hydrogenation process. At least, a thorough study of the step succession within the framework of the UBS HF approach under the described algorithm gives no chance for such opportunity. At the same time, the approach logically completes the $C_{60}H_{2k}$ series by the single isomer $C_{60}H_{60}$ of $I_h$ symmetry.

### 4.2. Activity of fluorination and hydrogenation reactions

Stepwise $C_{60}H_{2k}$ reactions activity can be characterized by the coupling energy $E_{cpl}$ that is needed for the addition of each next pair of atoms to the fullerene cage. Fig. 3 presents the quantity as a function of the step number $E_{cpl}(k)$ for both hydrogenation and fluorination processes. As seen from the figure, the functions behave quite similarly in both cases: starting at large absolute values at small $k$, $E_{cpl}$ gradually decreases when the step number increases and changes sign at $k\sim25$-26. At these steps any of both reactions becomes energetically non-profitable. The feature well explains the absence of either high hydrogenates or high fluorinates at $k >25$ among the products obtained.

Molecular chemical susceptibility $N_D$ is the other characteristic quantifier. The quantity is shown in the top part of Fig.3. As seen from the figure, the $N_D(k)$ functions are practically identical for both families gradually decreasing at higher $k$ and approaching zero at $k\sim20$-24. Therefore, decreasing $E_{cpl}$ by absolute value correlates with decreasing molecular chemical susceptibility $N_D$, or, by other words, with working out the pool of effectively unpaired electrons, which results in a considerable lowering of the reaction activity when $k$ changes from 18-20 to 25-26. According to both characteristics, the reaction is terminated at $k >25$. Important to note, that in spite of obvious obstacles for obtaining, $k$-high products might be abandoned among the final products. This is due to accumulative character of the reaction until the next addition of the atom pair is still energetically favorable. However, it is obvious that the accumulation efficacy will depend on the $E_{cpl}$ absolute value that is why more than twice difference in the $E_{cpl}$ absolute values for hydrogenates and fluorinates (see Fig.3) results in the termination of hydrogenation by $C_{60}H_{36}$ product while fluorination is completed by $C_{60}F_{48}$.

### 4.3. $C_{60}$ cage structure transformation during hydrogenation

Stepwise hydrogenation is followed by the gradual substitution of $sp^2$-configured carbon atoms by $sp^3$ ones. Since both valence angles between the corresponding C-C bonds and the bond lengths are noticeably different in the two cases, the structure of the fullerene cage becomes pronouncedly distorted. Fig.4 demonstrates the transformation of the cage structure in due course of hydrogenation exemplified by changes within a fixed set of C-C bonds.

A grey background diagram at each panel in Fig.4 presents the bond length distribution for the pristine $C_{60}$. As seen from the figure, the first steps of hydrogenation are followed by the elongation of C-C bonds that involve not only newly formed $sp^3$ atoms but some of $sp^2$ ones as well. This results in growing the total number of effectively unpaired electrons $N_D$ that is presented in details in Fig.5. The growing proceeds until $C_{60}H_{12}$ is formed after which the former effect is compensated by a gradual decreasing in the $N_D$ value caused by growing the number of $sp^3$ atoms.

Comparing the pristine diagram with those belonging to a current hydrogenate makes it possible to trace the fullerene cage structure changes. As might be naturally expected, the $sp^2$-$sp^3$ transformation causes the appearance of elongated C-C bonds, the number of which increases when hydrogenation proceeds. To keep the cage structure closed, this effect as well as changes in valence angles should be compensated. At the level of bonds this compensation causes squeezing a major part of pristine bonds, both long and short, therewith the squeezing effect grows when hydrogenation proceeds. This is particularly seen in the right-hand panels of Fig.4 related to $k$-high hydrogenates where a number of extremely short C-C bonds of 1.325-1.320 Å in lengths are observed. These six bonds are the only ones that remain unsaturated in $C_{60}H_{48}$ species and carry information about the pristine structure. Once each of them has been saturated when hydrogenation proceeds from $C_{60}H_{48}$ to $C_{60}H_{60}$, a new non-stressed cage of $I_h$ symmetry is formed. Important to note that in spite of the fact that all carbon atoms become $sp^3$-configured and should be considered as fully identical, the short-and-long bond pattern of pristine $C_{60}$ is kept for $C_{60}H_{60}$ as well, whilst with less difference between the two kinds of bonds. Moreover, previously short (long) bonds of $C_{60}$ keep their character in $C_{60}H_{60}$. Similar $sp^2$-$sp^3$ transformation of the cage was observed under fluorination from $C_{60}$ to $C_{60}F_{60}$ [4].

## 5. Conclusion

The reaction of fullerene $C_{60}$ with molecular hydrogen was studied using unrestricted broken symmetry HF SCF semiempirical approach (UBS HF version of the AM1 technique of the CLUSTER-Z1 codes). The calculations were focused on sequential addition of hydrogen molecules to the fullerene cage. A complete family of species $C_{60}H_{2k}$ ($k$=1,..., 30) was produced. Based on the effectively-non-paired-electron concept of the selectivity of the fullerene molecule chemical activity as well as on a suggested methodology of computational synthesis of fullerene derivatives [1-3], the synthesis of hydrogenates was performed as a series of predicted sequential steps. The preferred binding sites for sequential additions were selected by the largest value of atomic chemical susceptibility that is quantified by the effectively unpaired electron fraction $N_{DA}$ on the considered atom $A$ calculated at each step. This approach allowed for overcoming the difficulty connecting with many-fold isomorphism of the species as well as for releasing the computational scheme from the contiguous route of sequential additions. Added by the least total energy preference, the synthetic methodology allowed for obtaining a single isomer for any of $C_{60}H_{2k}$ ($k$=1,..., 30) species. The obtained results revealed a structural parallelism between $C_{60}$-hydrogenates and $C_{60}$-fluorinates [4]. However much weaker interaction of hydrogen molecules with the fullerene cage in comparison with that of fluorine molecules causes a significant difference in the species production limiting the hydrogenate family by $C_{60}H_{36}$ while $C_{60}F_{48}$ is the most abundant among fluorinates. The finding is well consistent with experimental picture of the $C_{60}$-hydrogetate-fluorinate parallelism, complete in general while different in details. A good fitting of the data to experimental findings convincingly proves a creative role of the suggested synthetic methodology in considering fullerene-involved addition reactions of different kinds.

**Chart 1    Hydrogenation algorithm in action**

| H0-C60 | | H2 (22, 33) | | H4 (60, 57) | | H6 (40, 54) | | H8 (55, 38) | |
|---|---|---|---|---|---|---|---|---|---|
| *Atom number* | *NDA* | *Atom number* | *NDA* | *Atom number* | *NDA* | *Atom number* | *NDA* | *Atom number* | *NDA* |
| **22** | **0.27101** | 5 | 0.29052 | 31 | 0.30726 | **55** | **0.35102** | 48 | 0.34871 |
| **33** | **0.27102** | **60** | **0.29051** | 34 | 0.30477 | 24 | 0.33194 | **52** | **0.33261** |
| … | | 3 | 0.29040 | **40** | **0.30247** | 31 | 0.29486 | 16 | 0.30018 |
| | | **57** | **0.29039** | **54** | **0.30142** | 37 | 0.29365 | 12 | 0.28547 |
| | | 42 | 0.26616 | 43 | 0.30044 | 34 | 0.28754 | 11 | 0.28333 |
| | | 9 | 0.2660 | … | | … | | … | |
| | | … | | *32* | *0.25687* | 23 | 0.27641 | *51* | *0.27866* |
| | | | | *35* | *0.25605* | … | | … | |
| | | | | … | | *38* | *0.26864* | 42 | 0.27144 |
| | | | | | | … | | … | |

**Isomer analysis by total energy,** *kcal/mol*

| | | | |
|---|---|---|---|
| **H6 (40, 54)** | **821.16** | **H8 (55, 38)** | **782.45** |
| H6 (31, 32) | 830.31 | H8 (24, 23) | 788.56 |
| H6 (34, 35) | 830.29 | | |
| H6 (31, 34) | 831.71 | | |

| H10 (52, 51) | | H12 (31, 32) | | H14 (24, 23) | | H16 (42, 48) | | H18 (58, 59) | |
|---|---|---|---|---|---|---|---|---|---|
| *Atom number* | *NDA* | *Atom number* | *NDA* | *Atom number* | *NDA* | *Atom number* | *NDA* | *Atom number* | *NDA* |
| 49 | 0.36041 | **24** | **0.40787** | 49 | 0.37686 | **58** | **0.41520** | 5 | **0.26506** |
| **31** | **0.35573** | 49 | 0.37738 | **42** | **0.35449** | 15 | 0.27992 | 4 | 0.26296 |
| 42 | 0.34512 | 42 | 0.34928 | *48* | *0.34201* | 46 | 0.27809 | 6 | 0.26036 |
| 48 | 0.34225 | 48 | 0.34036 | 29 | 0.29760 | 27 | 0.26521 | 53 | 0.25752 |
| 58 | 0.33829 | 29 | 0.30019 | 46 | 0.29412 | 5 | 0.26492 | 27 | 0.25628 |
| … | | … | | 47 | 0.29074 | … | | … | |
| 47 | 0.28156 | 47 | 0.28866 | … | | *59* | *0.25405* | *3* | *0.25100* |
| … | | … | | | | … | | … | |
| *32* | *0.27544* | *23* | *0.24824* | | | | | | |
| … | | … | | | | | | | |

**Isomer analysis by total energy,** *kcal/mol*

| | | | | | | | | | |
|---|---|---|---|---|---|---|---|---|---|
| H10 (48, 42) | 750.79 | H12 (49, 47) | 712.43 | **H14 (24, 23)** | **675.93** | H16 (49, 47) | 639.72 | **H18 (58, 59)** | **600.37** |
| **H10 (52, 51)** | **744. 05** | **H12 (31, 32)** | **711.31** | H14 (49, 47) | 678.98 | **H16 (42, 48)** | **637.59** | | |
| | | H12 (42, 48) | 713.14 | | | | | | |

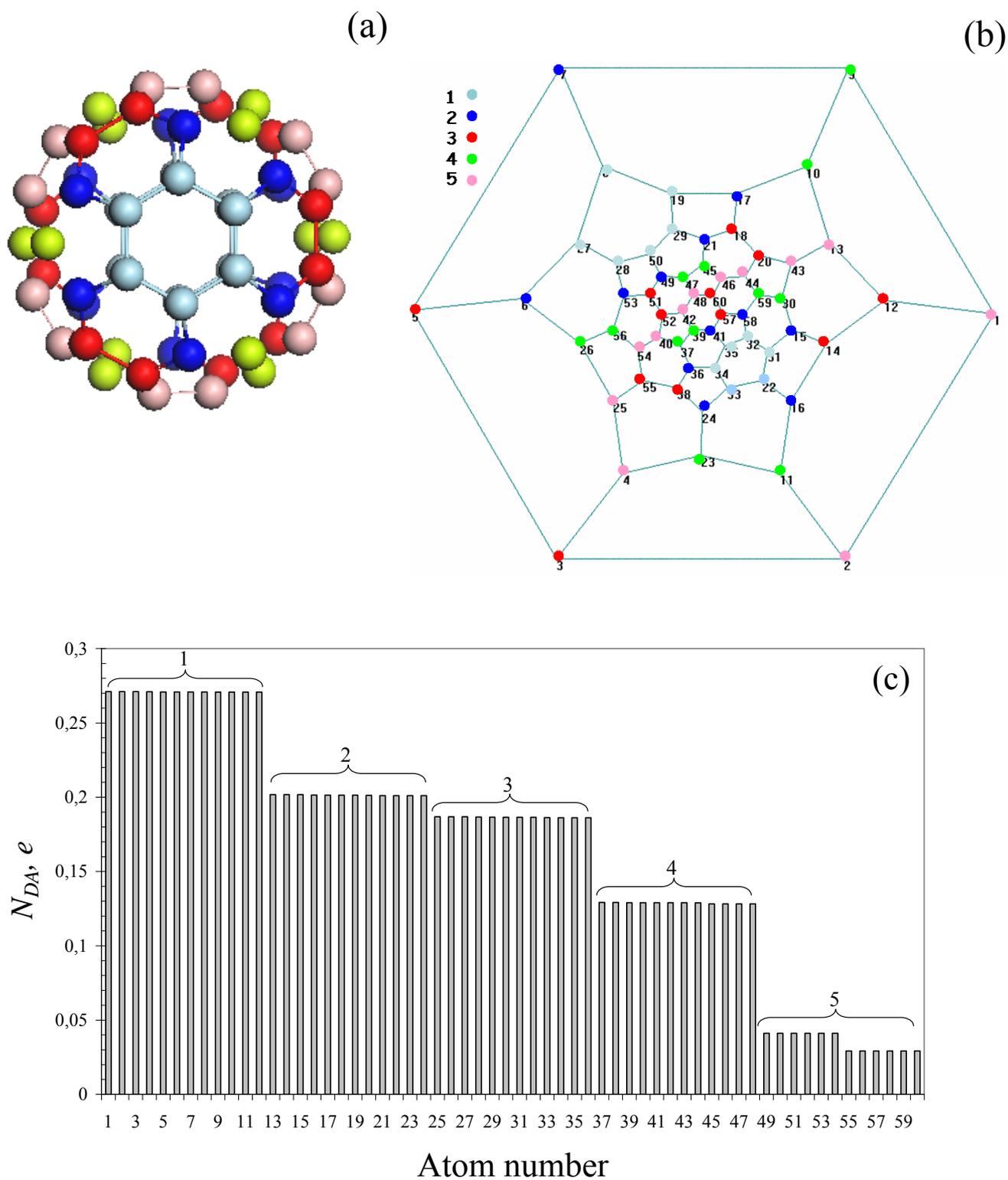

**Figure 1.** Chemical portrait of $C_{60}$ [1-3]. Color atom structure (*a*), Schlegel diagram (*b*), and distribution of atomic chemical susceptibility $N_{DA}$ over the molecule atoms (*c*). $N_{DA}$ data are aligned in A→Z manner. Color numeration in *b* corresponds to that of atom groups in *c*.

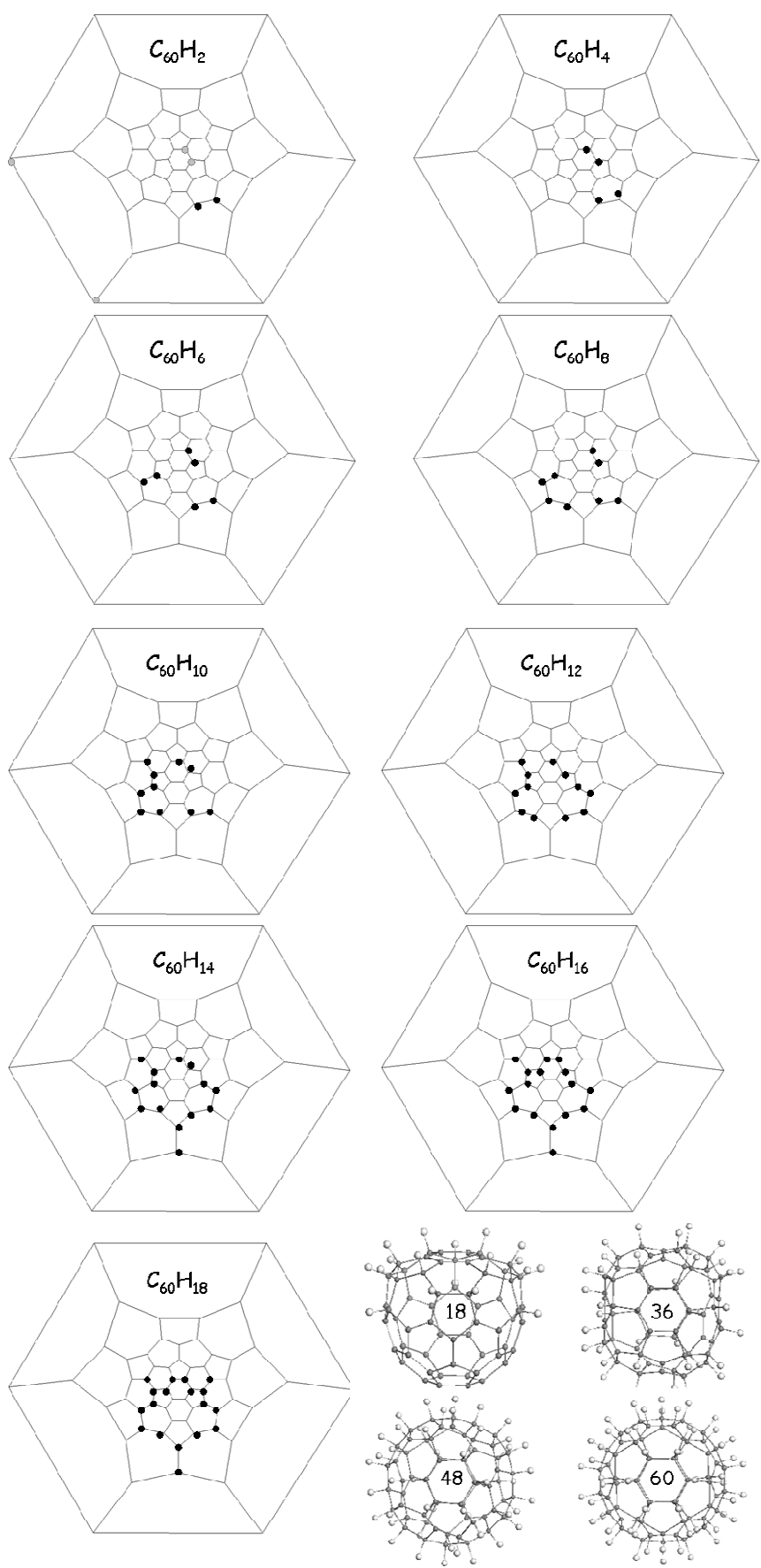

**Figure 2.** Schlegel diagrams of sequential steps of hydrogenation from $C_{60}H_2$ to $C_{60}H_{18}$ and atom structures of $C_{60}H_{18}$, $C_{60}H_{36}$, $C_{60}H_{48}$, and $C_{60}H_{60}$.

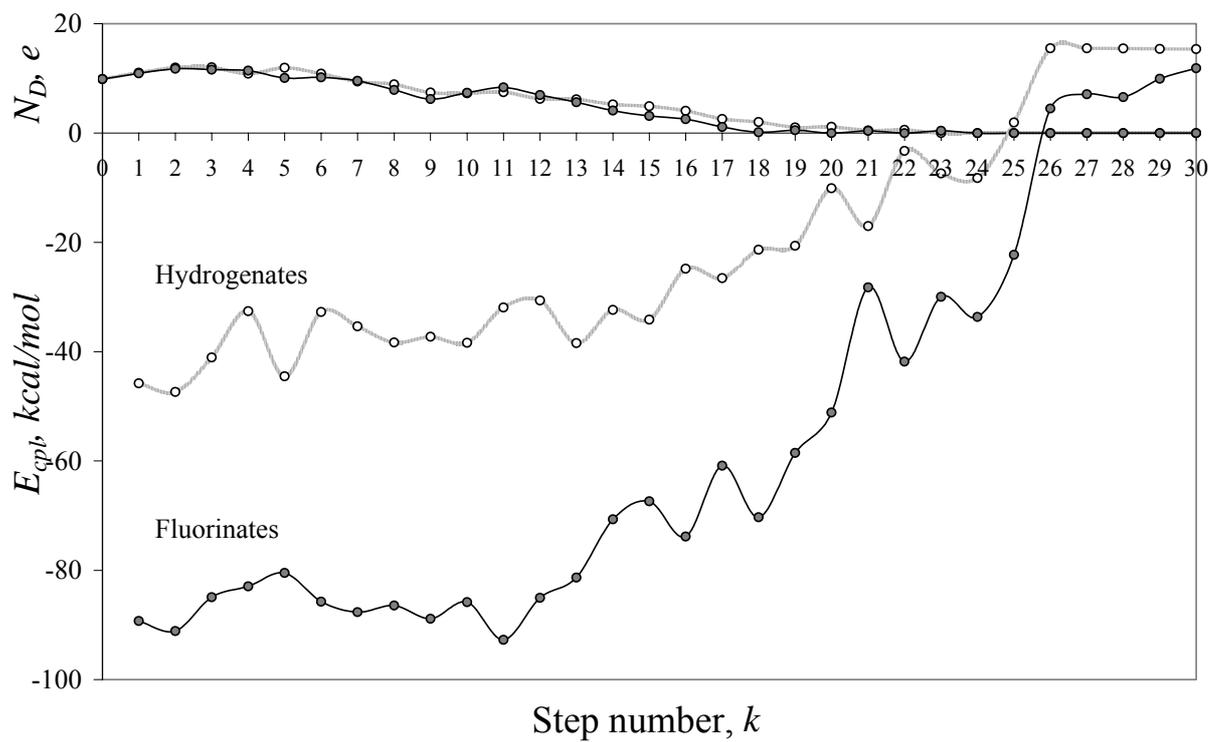

**Figure 3.** Coupling energy $E_{cpl}$ and molecular chemical susceptibility $N_D$ versus step number for $C_{60}$- hydrogenates (dotted curves with empty circles) and $C_{60}$- fluorinates (solid curves with filled circles).

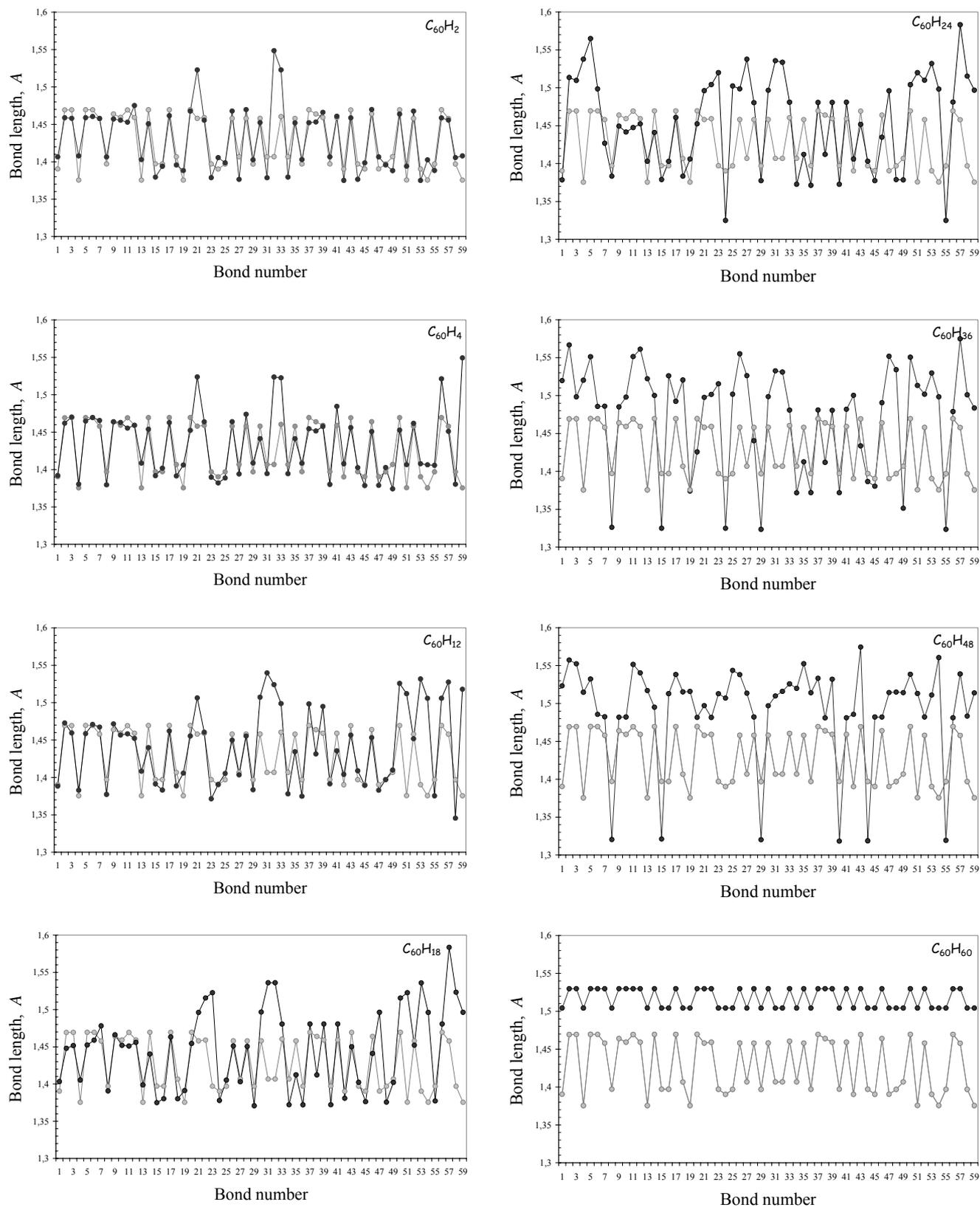

**Figure 4.** $sp^2$-$sp^3$ Transformation of the $C_{60}$ cage structure in due course of successive hydrogenation. Grey and black diagrams correspond to the bond length distribution related to the $C_{60}$ cage of pristine fullerene and its hydrogenates, respectively.

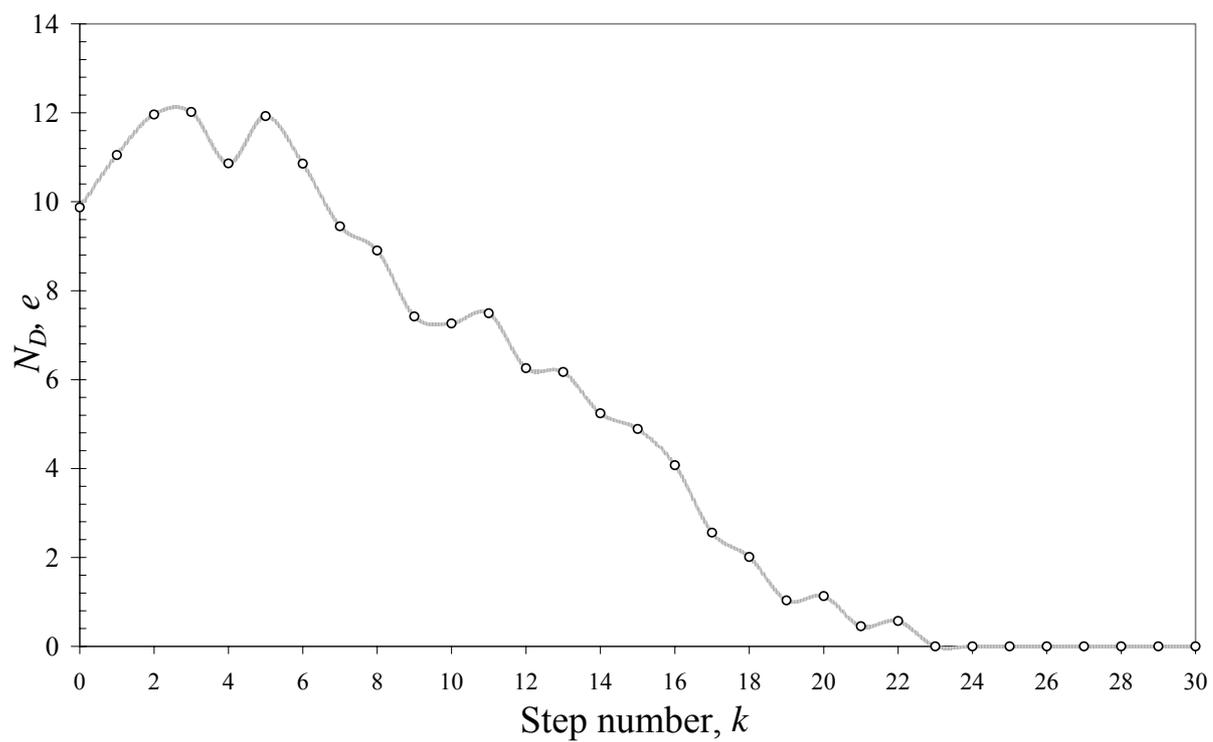

**Figure 5**. Molecular chemical susceptibility of $C_{60}H_{2k}$ hydrogenates versus step number.

Table 1. Geometric parameters, symmetry, and total energy of hydrogenated fullerene $C_{60}$ (UBS HF)

| | H2 | H4 | H 6 | H 8 | H10 | H 12 | H 14 | H 16 | H 18 | H 20 |
|---|---|---|---|---|---|---|---|---|---|---|
| $R(C^*-H)^a$, Å | 1,126 | 1,126 | 1,126-1,129 | 1,126-1,130 | 1,126-1,130 | 1,126-1,129 | 1,125-1,131 | 1,126-1,130 | 1,126-1,130 | 1,126-1,130 |
| $R(C^*-C^*)^b$, Å | 1,55 | 1,55, 1,52 | 1,54-1,50 | 1,54-1,50 | 1,54-1,50 | 1,54-1,50 | 1,54-1,50 | 1,54-1,50 | 1,54-1,50 | 1,54-1,50 |
| $\Delta H^d$, kcal/mol | 909,57 | 862,22 | 821,16 | 788,56 | 744,05 | 711,31 | 675,93 | 637,59 | 600,31 | 561,95 |
| Symmetry | $C_v$ | $C_s$ | $C_1$ | $C_1$ | $C_1$ | $C_1$ | $C_1$ | $C_s$ | $C_{3v}$ | $C_1$ |
| Atom pair | 22, 33 | 60, 57 | 40, 54 | 55, 38 | 52, 51 | 31, 32 | 24, 23 | 42, 48 | 58, 59 | 5, 3 |

| | H 22 | H 24 | H 26 | H 28 | H 30 | H 32 | H 34 | H 36 | H 38 | H 40 |
|---|---|---|---|---|---|---|---|---|---|---|
| $R(C^*-F)^a$, Å | 1,126-1,131 | 1,126-1,133 | 1,126-1,133 | 1,126-1,133 | 1,126-1,135 | 1,126-1,135 | 1,127-1,135 | 1,128-1,135 | 1,127-1,140 | 1,127-1,140 |
| $R(C^*-C^*)^b$, Å | 1,54-1,50 | 1,58-1,50 | 1,54-1,50 | 1,54-1,50 | 1,57-1,50 | 1,57-1,50 | 1,57-1,50 | 1,57-1,50 | 1,57-1,50 | 1,57-1,49 |
| $\Delta H^d$, kcal/mol | 530,03 | 499,37 | 460,93 | 428,56 | 394,43 | 369,60 | 343,06 | 321,72 | 301,07 | 290,98 |
| Symmetry | $C_1$ | $C_s$ | $C_s$ | $C_1$ | $C_1$ | $C_1$ | $C_s$ | $C_1$ | $C_1$ | $C_1$ |
| Atom pair | 6, 26 | 27, 28 | 13, 43 | 10, 17 | 8, 19 | 1, 2 | 12, 14 | 47, 49 | 50, 29 | 20, 18 |

| | H 42 | H 44 | H 46 | H 48 | H 50 | H 52 | H 54 | H 56 | H 58 | H 60 |
|---|---|---|---|---|---|---|---|---|---|---|
| $R(C^*-F)^a$, Å | 1,128-1,140 | 1,129-1,139 | 1,130-1,141 | 1,132-1,141 | 1,133-1,146 | 1,133-1,146 | 1,132-1,146 | 1,138-1,146 | 1,138-1,146 | 1,146 (1,112) |
| $R(C^*-C^*)^b$, Å | 1,57-1,49 | 1,57-1,50 | 1,57-1,49 | 1,56-1,48 | 1,55-1,48 | 1,55-1,49 | 1,55-1,48 | 1,54-1,49 | 1,53-1,50 | 1,53;1,50 |
| $\Delta H^c$, kcal/mol | 273,93 | 270,67 | 253,28 | 255,06 | 256,98 | 272,49 | 287,98 | 303,45 | 318,82 | 334,16 |
| Symmetry | $C_1$ | $C_1$ | $C_s$ | $C_1$ | $C_{2v}$ | $C_s$ | $C_{2v}$ | $C_{2h}$ | $C_{2v}$ | $I_h$ |
| Atom pair | 44, 46 | 35, 34 | 36, 37 | 25, 4 | 11, 16 | 53, 56 | 15, 30 | 21, 45 | 39, 41 | 7, 9 |

[a] C* mark cage atom to which hydrogen is added
[b] C*-C* mark a pristine short bond of the cage to which a pair of hydrogen atoms is added.
[c] $\Delta H$ is the heat of formation determined as $\Delta H = E_{tot} - \sum_A (E_{elec}^A + EHEAT^A)$. Here $E_{tot} = E_{elec} + E_{nuc}$, while $E_{elec}$ and $E_{nuc}$ are the electron and core energies. $E_{elec}^A$ and $EHEAT^A$ are electron energy and heat of formation of an isolated atom, respectively.